\documentclass[10pt,letterpaper]{article}

\usepackage{cogsci}
\usepackage{booktabs} 
\usepackage[table,xcdraw]{xcolor} 
\usepackage{graphicx}
\usepackage{amsmath}
\cogscifinalcopy 

\usepackage{pslatex}
\usepackage{apacite}
\usepackage{float} 

\title{Dimensions of Vulnerability in Visual Working Memory: An AI-Driven Approach to Perceptual Comparison}
 
\author{
    \large \bf Yuang Cao (12310901@mail.sustech.edu.cn)\textsuperscript{*} \\
    Department of Biomedical Engineering, Southern University of Science and Technology, Shenzhen, China 
    \AND 
    \large \bf Jiachen Zou (12210153@mail.sustech.edu.cn)\textsuperscript{*} \\
    Department of Biomedical Engineering, Southern University of Science and Technology, Shenzhen, China 
    \AND
    \large \bf Chen Wei (12150103@mail.sustech.edu.cn)\textsuperscript{\dag} \\
    Department of Biomedical Engineering, Southern University of Science and Technology, Shenzhen, China \\
    Department of Psychology, University of Birmingham, Birmingham, United Kingdom
    \AND
    \large \bf Quanying Liu (liuqy@sustech.edu.cn)\textsuperscript{\dag} \\
    Department of Biomedical Engineering, Southern University of Science and Technology, Shenzhen, China
}

\begin{document}

\maketitle

\begin{abstract}
Human memory exhibits significant vulnerability in cognitive tasks and daily life. Comparisons between visual working memory and new perceptual input (e.g., during cognitive tasks) can lead to unintended memory distortions. Previous studies have reported systematic memory distortions after perceptual comparison, but understanding how perceptual comparison affects memory distortions in real-world objects remains a challenge. Furthermore, identifying what visual features contribute to memory vulnerability presents a novel research question. 
Here, we propose a novel AI-driven framework that generates naturalistic visual stimuli grounded in behaviorally relevant object dimensions to elicit similarity-induced memory biases. We use two types of stimuli—image wheels created through dimension editing and dimension wheels generated by dimension activation values—in three visual working memory (VWM) experiments. These experiments assess memory distortions under three conditions: no perceptual comparison, perceptual comparison with image wheels, and perceptual comparison with dimension wheels. 
The results show that similar dimensions, like similar images, can also induce memory distortions. Specifically, visual dimensions are more prone to distortion than semantic dimensions, indicating that the object dimensions of naturalistic visual stimuli play a significant role in the vulnerability of memory.

\end{abstract}

\footnotetext[1]{* Equal contribution}
\footnotetext[2]{† Corresponding author.}
\section{Introduction}
Human visual working memory (VWM) serves as a critical cognitive interface, allowing us to temporarily retain and manipulate visual information to guide behavior~\cite{luck1997capacity,cowan2008differences,vogel2004neural}. Yet, this system exhibits marked vulnerability: even routine interactions with perceptual inputs—such as comparing a memorized object to a newly encountered one—can distort the original memory representation~\cite{fukuda2022working}. Such distortions challenge the fidelity of both visual and semantic memory features, such as misremembering the color of an item, or an eyewitness misremembering whether a suspect was holding a weapon or a repair tool. While prior work has established that perceptual comparisons induce retroactive biases in VWM (termed similarity-induced memory biases, SIMB), three critical questions remain unresolved: (1) How to construct naturalistic visual stimuli that systematically elicit dimension-specific memory distortions? (2) Can perceptual comparisons involving similar abstract dimension activations, like those with similar images, also induce memory distortions? and (3) Do visual and semantic dimensions exhibit differential levels of susceptibility to memory distortions?

Existing research on memory distortion has predominantly relied on simplified stimuli (e.g., colors, shapes) to isolate mechanistic principles~\cite{scotti2021visual,chunharas2022adaptive,saito2023comparing,saito2023perceptual,saito2024judgments}. Although these studies reveal that perceptual similarity amplifies memory biases, their conclusions are constrained by artificial experimental contexts. Real-world objects, in contrast, are defined by multidimensional and hierarchical attributes spanning low-level visual features (e.g., shape, texture) and high-level semantic properties (e.g., category, function). Recent evidence suggests that meaningfulness may enhance VWM capacity and stability~\cite{asp2021greater,sasin2023meaningful}, yet no study has compared the contribution of visual versus semantic dimensions to memory vulnerability. These gaps limit our understanding of how natural object representations interact with perceptual inputs to shape memory errors—a problem exacerbated by the lack of methods to precisely control and manipulate object dimensions in naturalistic stimuli.

The rise of AI-driven generative models offers a transformative solution. These models can synthesize naturalistic images, making them valuable tools for cognitive experiments~\cite{karras2019style,wei2024cocog1,wei2025synthesizing}. However, their application to memory research remains nascent, particularly in the context of disentangling and editing latent object dimensions. Here, we bridge this gap by employing a novel AI-driven generative model that generates natural visual stimuli through dimension manipulation~\cite{wei2024cocog2}. This approach allows us to construct stimulus gradients along both holistic perceptual variations and isolated dimension activations—a capability essential for addressing our three core questions.

\begin{figure*}[t] 
    \centering
    \includegraphics[width=1\textwidth]{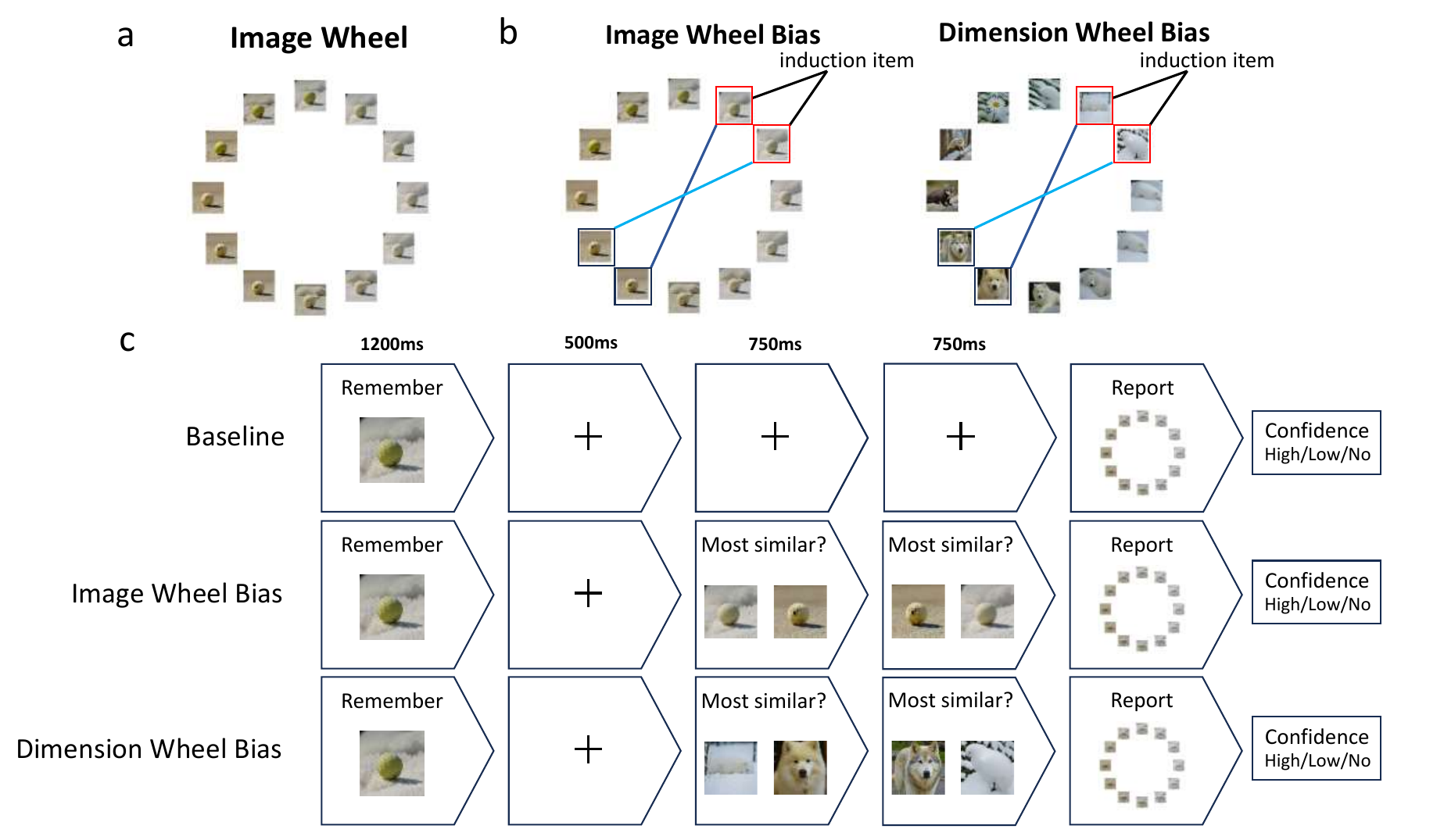} 
    \caption{\textbf{Experimental Design}.  
    (a) Image wheel: A circular arrangement of 12 images featuring gradual variation in characteristics. Participants were instructed to memorize the top image in the wheel (memory item).  
    (b) Bias induction: Image wheel bias trials used a pair of images include a induction item (red box, selected clockwise from the memory item randomly) and another image (blue box) for similarity judgments, while dimension wheel bias trials employed pairs from dimension wheels, where images in dimension wheels were typically perceived as dissimilar to the memory item.  
    (c) Experimental procedure: Three conditions (baseline, image wheel bias, dimension wheel bias). All conditions began with target memorization, followed by two similarity judgment phases (except baseline), and concluded with target recognition from the image wheel. Image pairs for similarity judgments were selected from image wheels in the image bias condition and from dimension wheels in the dimension bias condition.}
    \label{fig:overview}
\end{figure*}

In this study, we examined memory vulnerability under two conditions: perceptual comparison with image wheels and perceptual comparison with dimension wheels (Fig.~\ref{fig:overview}). Image wheels are generated by smoothly editing dimension activation values of an image, while dimension wheels are constructed from predefined activation values of dimensions in a latent space. These stimuli allow us to isolate the contributions of image similarity and dimension similarity to memory distortion during perceptual comparisons, as well as to disentangle the contributions of individual dimensions. We hypothesized that perceptual comparisons with both types of wheels would induce memory distortions, with visual dimensions exhibiting greater vulnerability than semantic dimensions. This hypothesis aligns with neurocognitive models positing that semantic features are better remembered thanks to schema-based representations and deeper processing~\cite{brady2016working,brady2022role,chung2023no}.

Our results confirm that similar object dimensions, whether manipulated via images or abstract dimension activations, can induce significant memory distortions. Additionally, visual dimensions showed markedly higher distortion susceptibility compared to semantic dimensions. These findings advance theoretical frameworks by demonstrating that memory vulnerability is not merely a function of perceptual similarity but is dimensionally structured. Methodologically, our AI-driven approach provides a blueprint for studying complex cognitive processes with naturalistic yet controlled stimuli, offering implications for AI-assisted experimental design and computational models of memory.

\begin{figure*}
    \centering
    \includegraphics[width=1\textwidth]{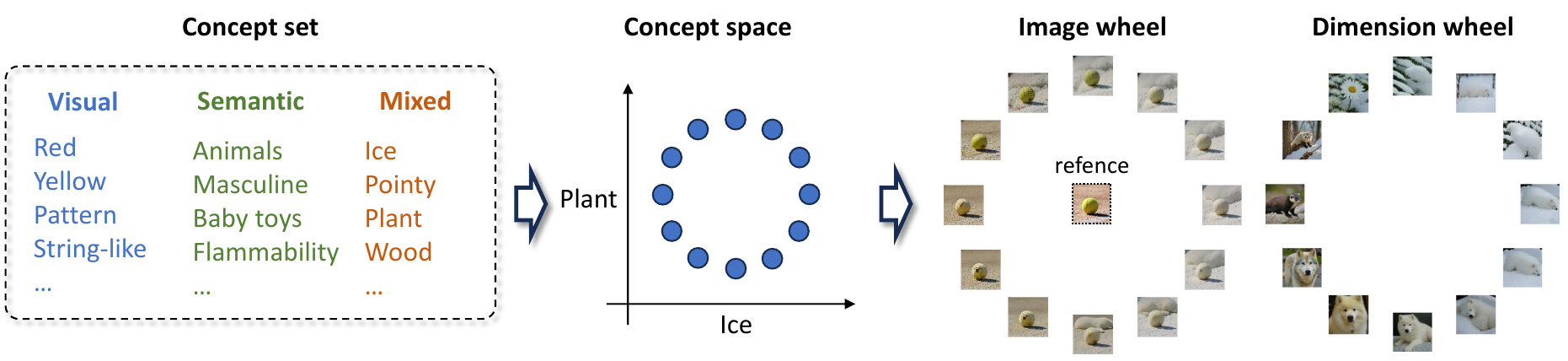} 
    \caption{\textbf{Wheel generation.}  The leftmost image shows the dimensions (with categories of visual, semantic, and mixed) that we selected during image generation. We created circles in the representation space of dimension-pairs as target dimension activation values, and then used an AI-driven generative model to generate wheels based on dimensions. For the image wheel, we generated wheels that are similar to the reference images but with different dimension activation values. For the dimension wheel, we used only the activation values in the dimension-pair space to generate wheels.}
    \label{fig:generation}
\end{figure*}

\section{Related Works}
\textbf{Similarity-Induced Memory Bias.} Studies have shown that when individuals compare a memorized stimulus to a similar perceptual input, the memory tends to shift towards the features of the new input, a phenomenon known as SIMB~\cite{fukuda2022working}. In SIMB experiments, participants were asked to remember a target stimulus and later reproduce it by selecting from a continuous wheel of stimuli. When a pair of probe stimuli was introduced during the retention period, the memory of the target was significantly biased toward the probe judged to be similar~\cite{saito2023comparing}. While SIMB has been observed with simple stimuli such as colors and shapes, its application to more complex, real-world objects remains understudied. Previous work primarily focused on low-level features, like color or shape~\cite{fukuda2022working,saito2023comparing,saito2023perceptual,saito2024judgments}, but real-world objects are defined by multiple features~\cite{hebart2020revealing,bracci2023understanding}, including both low-level visual attributes (e.g., shape, texture) and high-level semantic properties (e.g., function, category). This suggests that evaluating SIMB with dimension-based sythetic stimuli, is essential to understanding how image and dimension similarity contribute to memory distortions, as well as the respective contribution of visual and semantic dimensions. This is particularly relevant when studying objects in their natural context, where dimensionality and feature diversity play significant roles in the perception and memory of objects.

\textbf{Behavior-Based Object Dimensions.} Over the past few years, behavioral-based object dimensions have attracted considerable attention, especially in computer vision and cognitive science research. Previous work \cite{hebart2019things, hebart2020revealing, zheng2019revealing, muttenthaler2022vice} has gathered human similarity judgment data for a naturalistic dataset of 26,107 object images through extensive behavioral experiments. These studies have decoded the representation dimensions of images by either optimizing object-specific representations or leveraging deep neural networks (DNNs) to predict human behavior, thus reducing complex visual information into interpretable, low-dimensional object features. Notably, recent findings \cite{kramer2023features} suggest that these dimensions can be used to construct an object feature model capable of predicting image memorability. Collectively, these studies highlight the value of behavior-based object dimensions in understanding visual perception and memory.


\section{Dimension-Guided Wheel Generation}


We employed the Concept-based Controllable Generation model  from \cite{wei2024cocog1,wei2024cocog2} to generate visual stimulus wheels. The model adopts a two-stage generation strategy: first modeling $p(h|e)$ and subsequently modeling $p(x|h)$, where $e$ denotes conditioning parameters (dimension activation values or similarity), $h$ represents CLIP embeddings, and $x$ is the generated image. By incorporating training-free guidance during the modeling of $p(h|e)$, the model enhances generation flexibility and applicability. This framework allows effective guidance of visual stimulus generation through appropriate selection of conditioning parameters $e$ and differentiable loss functions $\ell(f_\phi(\cdot), \cdot)$ based on experimental objectives.

For the generation of image wheels, we randomly select an image from the THINGS dataset and perform dimension editing on it. We used the following loss functions to guide the generation process:

1. \textit{Dimensional Guidance}: We select two dimensions from the concept of the given image for editing. As shown in Fig. 1, we pre-designed the concept pair activation values for 12 images, arranging the activation values in a circular pattern within the representation plane defined by the concept pair axes. This guides the generation of a smooth transition of concepts across the image wheel. The corresponding loss function is:
\begin{equation}
\ell = \sum \| g(h)_i - c_i \|
\end{equation}
where $c_1$-$c_{12}$ constitute circular coordinates in the conceptual plane.

2. \textit{Smoothness Guidance}: To ensure the images on the image wheel maintain similarity and smooth transitions, we implemented:
\begin{equation}
\ell = \sum_{i,j \in S} \| h_i - h_j \|
\end{equation}
where $\{i,j\}$ denotes indices of neighboring images in the wheel.

3. \textit{CLIP Guidance}: To preserve latent feature similarity with the given image, we used the following loss function:
\begin{equation}
\ell = \| h - \bar{h} \|
\end{equation}
where $\bar{h}$ represents the CLIP embedding of the source image.

4. \textit{Pixel Guidance}: We control the pixel-level similarity of the images by introducing img2img~\cite{meng2021sdedit}, which uses a noisy version of the given image as the starting point for generation.

For the generation of \emph{dimension wheels} , we only applied dimension guidance. For each \emph{image wheel}, we aimed to generate a dimension wheel by setting the predefined dimension pair activation values as the target. Since there are no restrictions on image similarity, the dimension wheels have lower visual similarity but maintain consistent dimensional similarity. This effectively separates the effects of visual similarity and dimension similarity on memory distortion.


\begin{table*}[t]
\centering
\renewcommand{\arraystretch}{1.4} 
\begin{tabular}{|l|l|l|}
\hline
\rowcolor{gray!40} \textbf{Semantic} & \textbf{Visual} & \textbf{Mixed} \\
\hline
\rowcolor{gray!10} aquatic activities/sea & balls/spherical & decorative/gold \\
baby toys/kid & rope/twine & green/vegetable \\
\rowcolor{gray!10} baked food/healthy & circular/discs & home tools/silver \\
body accessories/hair & groups of small objects/multicolored & ice/white \\
\rowcolor{gray!10} body part-related/brittleness & many colors/multicolored & natural minerals/crystal \\
containers for liquids/beverage & pattern/texture & paper-like/paper \\
\rowcolor{gray!10} cotton clothing/clothing & pole/sticks & plant/green \\
face accessories/eyes & red/shininess & pointy/tools \\
\rowcolor{gray!10} flammability/fire & string-like/netting & weaponry/silver \\
flying to not/flight & yellow/gold & wearable/black \\
\rowcolor{gray!10} ground animals/mammal & & wood/brown \\
household furniture/furniture & & \\
\rowcolor{gray!10} masculine/limbs & & \\
medical supplies/absorbency & & \\
\rowcolor{gray!10} music instruments/instruments & & \\
natural resources/terrestrial & & \\
\rowcolor{gray!10} old technology/used & & \\
outdoor objects/outdoor & & \\
\rowcolor{gray!10} recreational instruments/sport & & \\
things with wheels/transportation & & \\
\rowcolor{gray!10} wheeled vehicles/transportation & & \\
\hline
\end{tabular}
\caption{\textbf{Dimension Categorization.} Classify the 42-dimensional behavior-based object dimensions across semantic, visual, and mixed dimensions.}
\label{tab:category}
\end{table*}

\section{Image Wheels Induction Experiment}

\begin{figure}
    \centering
    \includegraphics[width=0.5\textwidth]{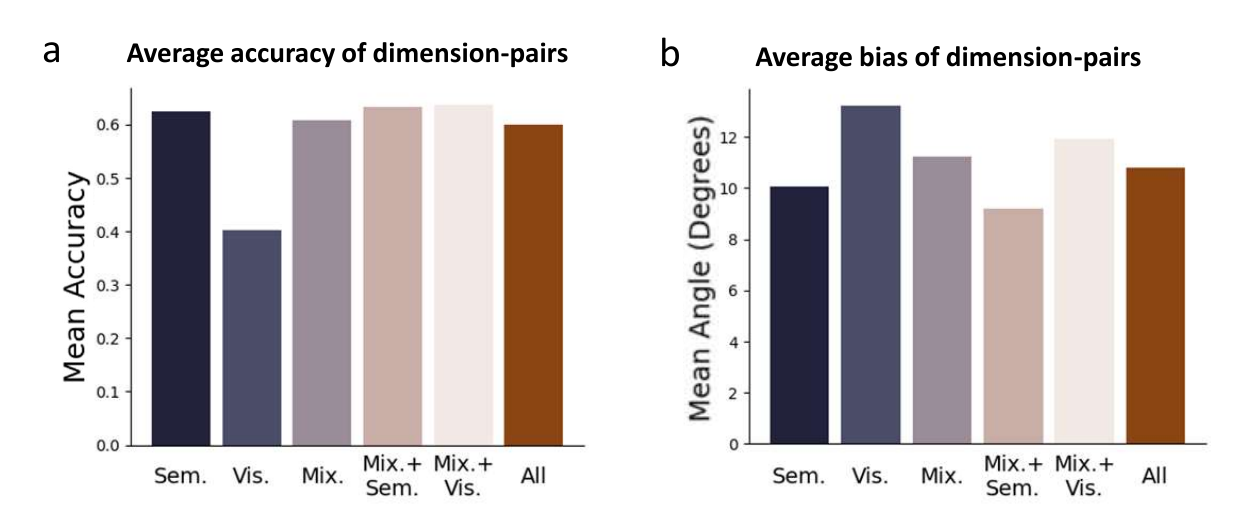} 
    \caption{\textbf{Experimental results of the Image Wheels Induction Experiment (N=100).} 
    (a) Mean accuracy of memory performance across dimension-pair categories. 
    (b) Mean bias scores for dimension-pairs. \\
    \textbf{Sem.} = Semantic, \textbf{Vis.} = Visual, \textbf{Mix.} = Mixed.}
    \label{fig:exp-img}
\end{figure}

\subsection{Stimuli and Task Design}

Image wheels were constructed using an AI-driven generative model that smoothly interpolated latent dimension activations to produce naturalistic object variations. Each wheel represented a circular manifold of images, where angular positions corresponded to incremental changes along predefined object dimensions. A memory item was randomly selected from an image wheel and presented for 1,200 ms. After a 1,500 ms retention interval, the picture wheel appeared, and the subject reported which picture in the wheel was the original memory item. 

During the induction phase, two items were simultaneously displayed: an \textit{induction item} and a \textit{dissimilar item}. The induction item was sampled from a 60° arc along a target dimension direction relative to the memory item, while the dissimilar item occupied the position 180° opposite to the induction item. This design ensured that the induction item shared perceptual similarity with the memory item along the manipulated dimension, whereas the dissimilar item served as a control. Participants performed a same-different judgment task on these pairs for 1,500 ms. After a 500 ms retention interval, the picture wheel appeared, and the subject reported which picture in the wheel was the original memory item. The interstimulus interval is determined based on the results of the pre-test, in order to ensure that participants experience comparable task loads across different experimental conditions, thereby minimizing the potential interference of temporal variables on memory performance.


A total of 100 participants completed the experiment on the Brain Island platform. Each participant performed 120 trials, including 30 no-induction baseline trials and 90 induction memory trials. The experimental stimuli were categorized into three groups based on feature dimensions: \textit{visual} (e.g., shape-color), \textit{semantic} (e.g., tool-animal), and \textit{mixed} (e.g., color-category). The mixed condition integrates both visual and semantic information dimensions, and we collected data from this condition to ensure the completeness of the study. However, the primary aim of this research was to compare the memory vulnerability differences between visual and semantic dimensions, so we did not conduct an in-depth analysis of the mixed condition. The results from the mixed condition were only used as supplementary information to help us gain a more comprehensive understanding of the potential impact of visual and semantic information integration on memory. The accuracy of the no-induction experiment reflects the participant’s ability to remember the images. If a participant's overall accuracy in the no-induction experiment is below 40\%, we do not accept their experimental results. In the induction experiment, the participant undergoes two induction trials. We consider the participant's final selection results in the image wheel to be valid only if they select the \textit{induction item} in both induction trials, as this indicates their memory of the image is accurate. Additionally, at the end of each experimental trial, participants are asked to rate their confidence in their selection. We excluded low-confidence trials because they are less likely to reflect the participant’s true memory and are more likely to be based on uncertainty or guessing. Excluding these trials helps ensure that the memory distortions we observe are due to actual memory effects rather than guesses.

\subsection{Behavioral Measures and Analysis} 
Memory performance was quantified using two metrics: (1) \textit{accuracy}, defined as the proportion of correct probe identifications, and (2) \textit{bias scores}, calculated as the angular deviation between participants’ memory reports and the original memory item, with higher scores indicating stronger distortion toward the induction direction. The bias score is calculated using the following formula:

\begin{equation}
\text{Bias Score} = |\theta_{\text{Report}} - \theta_{\text{Target}}|
\end{equation}

where \(\theta\) represents the angular coordinate. For example, if the angle deviation between the memory report and the original item is 15°, the bias score would be 15°.

\subsection{Results}
The experiment revealed systematic differences in memory vulnerability across dimensions (Figure 3b–c). For accuracy, visual dimensions (0.372) and mixed+visual combinations (0.484) showed significantly lower performance compared to semantic dimensions (0.503) and mixed+semantic combinations (0.498). Bias scores further highlighted this asymmetry: visual dimensions exhibited the strongest distortion (13.210°), followed by mixed+visual pairs (11.930°), whereas semantic dimensions (10.053°) and mixed+semantic pairs (9.194°) showed comparatively weaker biases.  

\subsection{Interpretation} 
These results demonstrate that perceptual comparisons with image similarity robustly induce retroactive memory distortions, specifically reflected in the tendency of participants to recall memory content that is biased toward the induced items with similar characteristics, with visual dimensions being disproportionately vulnerable. The gradient structure of the image wheels allowed us to isolate dimension-specific interference, confirming that even naturalistic, multidimensional objects are subject to similarity-driven biases. The weaker distortion in semantic dimensions aligns with theories positing that schematic or categorical representations stabilize memory traces against interference.

\section{Dimension Wheels Induction Experiment} 

\begin{figure}[t]
    \centering
    \includegraphics[width=0.5\textwidth]{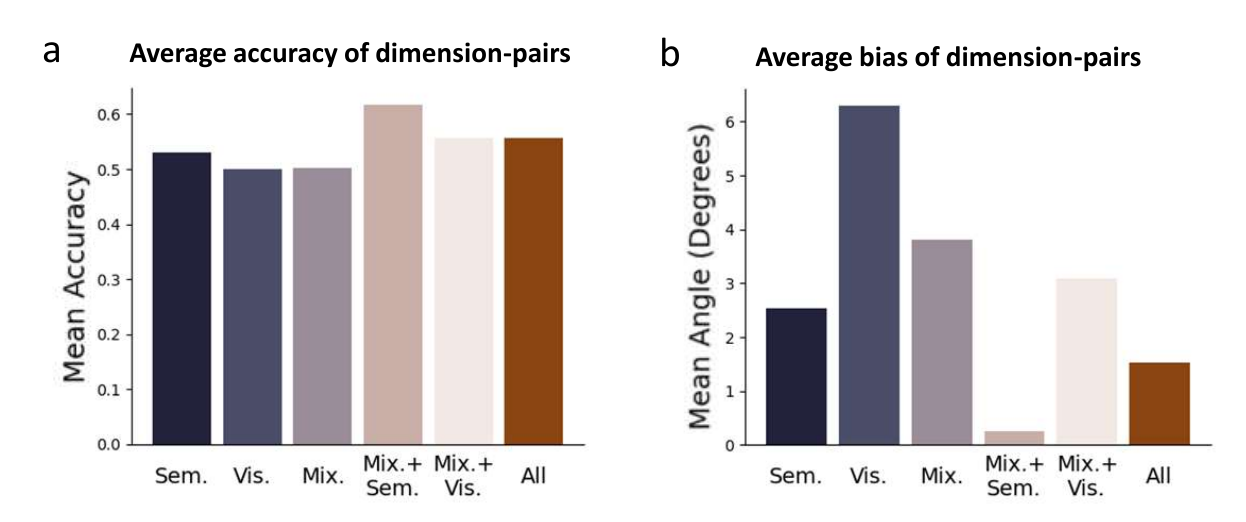} 
    \caption{\textbf{Experimental results of the Dimension Wheels Induction Experiment(N=146).} (a) Mean accuracy of memory performance across dimension-pair categories. (b) Mean bias scores for dimension-pairs.}
    \label{fig:exp-dim}
\end{figure}

\subsection{Stimuli and Task Design}

Dimension wheels were constructed by embedding predefined activation values of visual or semantic dimensions into a latent space using an AI-driven generative model. Unlike image wheels, which interpolate between holistic perceptual variations, dimension wheels explicitly manipulate isolated dimensions. 




A total of 146 participants took part in this experiment on the Brain Island platform. The experimental procedure, total number of trials, and selection criteria for the dimension wheel experiment were the same as those for the image wheel experiment, with the only change being that all induction items were drawn from the dimension wheel.

\subsection{Behavioral Measures and Analysis} 
As in the Image Wheels experiment, memory performance was assessed via \textit{accuracy} (proportion of correct responses) and \textit{bias scores} (angular deviation toward the induction item). 

\subsection{Results}
The experiment revealed a pronounced hierarchy in dimension-specific memory vulnerability (Fig.~\ref{fig:exp-dim}). For accuracy, visual dimensions (mean accuracy = 0.500) and mixed+visual combinations (0.557) underperformed relative to semantic dimensions (0.530) and mixed+semantic pairs (0.617). Bias scores further emphasized this pattern: visual dimensions exhibited the strongest distortion (mean bias = 6.290°), followed by mixed pairs (3.805°) and mixed+visual pairs (3.090°), while semantic dimensions (2.530°) and mixed+semantic pairs (0.254°) showed minimal biases. 

\subsection{Interpretation} 
These results confirm that memory distortions can be induced even by abstract dimension activations, independent of holistic perceptual similarity. The significantly weaker biases in semantic dimensions compared to visual dimensions, suggest that semantic features benefit from conceptual hierarchies or schema-based stabilization. 

In addition, although the overall memory distortion induced by the dimension wheel (accuarcy = 0.556, bias score = 1.520°) was smaller than that of the picture wheel (accuarcy = 0.470, bias score = 10.780°), the difference between the semantic dimension and the visual dimension was significantly larger in the dimension wheel. By decoupling dimension activations from perceptual context, this experiment advances computational models of VWM, highlighting the need to account for both feature-specific and integrative similarity mechanisms.


\section{Discussion}

Our findings provide novel insights into the dimensional architecture of vulnerability in visual working memory. By leveraging AI-driven generative models to disentangle and manipulate object dimensions in naturalistic stimuli, we addressed three critical gaps in the literature: (1) the construction of dimension-specific naturalistic memory probes, (2) the role of abstract dimension activations in inducing memory biases, and (3) the differential susceptibility of visual versus semantic features to retroactive interference.

\textbf{Dimension-Specific Vulnerability in Memory.}
As predicted, both holistic (image wheels) and dimension-specific (dimension wheels) comparisons distorted memory. Visual dimensions showed greater distortion than semantic ones, supporting models that link semantic memory to deeper, schema-based processing~\cite{brady2016working,chung2023no}. In contrast, early-stage visual features are more prone to interference from similar stimuli.
This visual-semantic asymmetry challenges purely perceptual accounts of similarity-induced memory biases~\cite{scotti2021visual,saito2023perceptual}, suggesting that memory errors are also dimensionally organized. This has practical implications—for instance, enhancing semantic context could reduce memory distortions in settings like eyewitness testimony or AI-assisted tasks.

\textbf{AI-Driven Methods for Cognitive Research.}
Our approach demonstrates how generative models can isolate and manipulate object dimensions in complex, realistic stimuli. Unlike traditional methods using simple shapes, our framework preserves naturalistic detail while enabling precise control. This allows for studying feature interactions in memory, and potentially in other areas like attention or decision-making.
The distinction between image and dimension wheels further validates this method. While image wheels caused greater overall distortion, dimension wheels revealed a clearer divergence between visual and semantic memory errors. These findings point to the need for memory models to account for multidimensional similarity, not just perceptual distance.

\textbf{Limitations and Future Directions.}
While our AI-driven approach advances stimulus control, several limitations warrant attention. First, our stimuli, though naturalistic, were constrained to predefined latent dimensions; real-world objects may exhibit emergent features not captured by current generative models. Second, our experiments focused on static comparisons, whereas dynamic interactions (e.g., sequential object manipulations) might reveal additional mechanisms of memory distortion. Finally, individual differences in cognitive style or expertise—factors known to influence schema formation—were not explored but could further explain variability in dimensional vulnerability.




\section{Acknowledgments}

This work was supported by the National Natural Science Foundation of China (62472206), Shenzhen Science and Technology Innovation Committee (2022410129, KJZD20230923115221044, KCXFZ20201221173400001), Guangdong Provincial Key Laboratory of Advanced Biomaterials (2022B1212010003), and the Center for Computational Science and Engineering at Southern University of Science and Technology.







\bibliographystyle{apacite}

\setlength{\bibleftmargin}{.125in}
\setlength{\bibindent}{-\bibleftmargin}

\bibliography{CogSci_Template}

\end{document}